\documentstyle[twocolumn,aps]{revtex}

\begin{document}
\draft              
\twocolumn[\hsize\textwidth\columnwidth\hsize\csname @twocolumnfalse\endcsname

\title{Level statistics and localization in a 2D quantum percolation problem}

\author{M.~Letz$^{a,b}$ and K.~Ziegler$^{a,c}$
}

\address{
$^a$Max--Planck--Institut f\"ur  Physik Komplexer Systeme, Au{\ss}enstelle
Stuttgart, Heisenbergstr. 1, D-70506 Stuttgart, Germany}
\address{$^b$Queen's University, Department of Physics, Kingston, Ontario,
  K7L 3N6 
  Canada} 
\address{$^c$Institut f\"ur Physik, Universit\"at Augsburg, D-86135 Augsburg,
Germany}

\date{\today}
\maketitle

\begin{abstract}
A two dimensional model for quantum percolation with variable tunneling range
is studied. For this purpose the Lifshitz model is considered where the 
disorder enters the Hamiltonian via the nondiagonal elements. 
We employ a numerical method to analyze the level statistics of
this model. It turns out that the level repulsion is strongest around the
percolation threshold. As we go away from the maximum level repulsion a
crossover from a GOE type behavior to a Poisson like distribution is
indicated. The localization properties are
calculated by using the sensitivity to boundary conditions and we find a 
strong crossover from localized to delocalized states. 

\end{abstract}

]
\vskip 0.5cm 
\narrowtext

The statistics of energy levels in complicated quantum systems has been a
subject of research for several decades. It started with the study of energy
levels in atomic nuclei \cite{Wigner1}, then the statistics of
electronic 
states in atoms was investigated \cite{Rosenzweig}, and more recently the
statistics of electrons in quantum dots (also known as ``artificial'' atoms)
\cite{efetov}. 
A surprising result of most of these approaches was that the statistics of
energy levels is quite universal regardless of the specific system: systems
can be classified according to their symmetry properties as orthogonal,
unitary and symplectic. These universality classes represent strong
correlations between the energy levels due to level repulsion.
This is indicated, e.g., by the level
spacing distribution $P(s)$ which goes like $s^\beta$ for small $s$. The
exponent $\beta$ (=1,2,4) characterizes the universality class
\cite{metha91}. In contrast
to this repulsive behavior (Wigner--Dyson distribution)
the assumption of statistically independent energy levels would lead to a 
Poisson distribution $P(s)=e^{-s}$. The
correlation in nuclei or atoms is so strong because their corresponding states
have usually a large overlap, except perhaps for the ground state. The
situation is less clear if we consider a macroscopic system of atoms like in
solid state physics where disorder can be present. 
Depending on the latter there are extended electronic states
and also states which are localized in space due to
Anderson localization \cite{efetov}. 
In terms of level statistics the localized states are expected to obey a
Poisson distribution, whereas the extended states are expected to have strong
level repulsion characterized by a Wigner--Dyson distribution.
Since there is a metal-insulator transition from extended to localized states
driven by disorder (Anderson transition), it is natural to study
this in terms of the statistics of energy levels. 

Approaches to the role of level statistics for the Anderson transition 
\cite{evangelou94} were presented by several authors, and a model for 
quantum percolation for nearest neighbor transfer
was investigated in Ref. \cite{berkovits96}. Recently 
it was found \cite{aronov9495} that the divergence of the localization length
$\xi$ at the metal--insulator transition leads to a deviation from the
Wigner--Dyson statistics for $\xi > L$, $L$ being the system size. 
In particular it was found that the decay of the level spacing density is 
weaker than that of the Wigner--Dyson statistics. This is also found in our 
investigation.

The purpose of this paper is to study a quantum percolation model where the
transfer is not only between nearest neighbors but where the transfer rate
decays exponentially with distance.
Moreover, in contrast to Anderson's model for localization we study disorder 
in the off-diagonal part of the Hamiltonian. This is also known as Lifshitz 
type of disorder.
This model is motivated by physical systems.  
One example is a two-dimensional array of quantum dots
\cite{duruoez95}. Our model is also motivated by the analogous picture of 
variable range hopping in solids \cite{mott74}. 
A third example for this model is the class of low doped high--T$_c$ cuprates.
Here 
the charge carriers are holes in two--dimensional CuO$_2$ layers. It seems
that disorder and phase separation play an essential role in these systems
\cite{sighizh,emery96}. The two--dimensional copper oxide plane is
separated into hole rich conducting and magnetically correlated  (insulating)
areas. A possible origin for phase separation are polaronic states that are
discussed in Refs. \cite{klemm,gooding}. Therefore our model can describe a
transition or at least a crossover from strongly insulating (localized)
states to states with infinite or at least very large localization
length. This picture can be applied to the physics of the normal state in the
low--doped high--T$_c$ materials.
Indeed, conductivity measurements by Chen et al. \cite{chen95} of low doped
La$_{2-x}$Sr$_{x}$CuO$_{4}$ with $x\approx $ 0.002 have shown that
for temperatures below 50 K the transport properties are governed by a
hopping type conduction. Also earlier measurements by Keimer et al.
\cite{keimer92}
for a sample with doping concentrations of $x\approx $ 0.04 found a
conductivity of hopping type near localization below 20 K.  

Our model for quantum percolation 
corresponds to the Hamiltonian
\begin{equation}
\label{eq1}
H = \sum_{i,j }( t_{ij} a_i^{\dagger} a_{j} + h.c. ) \; ,
\end{equation}
with the following off--diagonal (hopping) matrix elements
\begin{equation}
\label{eq2}
t_{ij} = \left\{ 
\begin{array}{lr}
t \exp(-\alpha(r_{ij} - r_0)) & \mbox{ for } r_{ij} > r_0 \\
t                            & \mbox{ for } r_{ij} \leq r_0
\end{array} 
\right. \; .
\end{equation}
The lattice sites $i$ and $j$ can be randomly occupied with 
quasiparticles in Wannier states, leading to random hopping elements
$t_{ij}$.
This type of randomness is also known as Lifshitz type disorder 
\cite{lifshitz65}.
The exponential decay of the localized wave functions leads to an
exponentially decay of the hopping rate with distance on the
inverse decay length $\alpha$. The spatial extension of the localized 
states, e.g., given by the size of a polaron, is expressed with $r_0$. 
A hopping matrix element $t_{ij}$ is non--zero only if the sites $i$
and $j$ of the 2D lattice are both occupied by localized states.

An advantage of the long range hopping of our model, at least for small 
enough $\alpha$, is the fact that the density of states is smoothed out in 
contrast to the sharp peaks found for nearest neighbor transfer 
\cite{berkovits96}. The smooth density of states is 
easier to analyze with the methods of random matrix theory. The density of 
states is shown in fig. \ref{fig3}. It shows a broad peak near the lowest 
eigenvalue. This peak is due to the
two--dimensional nature of the system. For infinite $\alpha$ and
all lattice sites occupied only a nearest neighbor transfer
remains. In this case the density of states is the elliptical integral
with the logarithmic singularity at the center. For finite $\alpha$
also next nearest neighbor and further transfers are included. This
shifts the peak in the density of states to the lower band edge. The
density of states drawn in fig. \ref{fig3} shows a remnant of this peak.

The numerical calculation is performed as follows. 
The $N$ (typically $N = 400$, $N < L^2$) localized states are randomly
chosen with probability $c$ ($c=N/L^2$) on an $L\times L$ square lattice 
with lattice constant $a$. In this procedure periodic boundary conditions are
used. The coordinates of the localized states are distributed randomly
while multiple occupation is prohibited.

For example, for $r_0 = a $ the classical (bond) percolation threshold is 
near
$c \approx 0.5$. For any combination of pairs the off-diagonal
elements have to be computed for the Hamiltonian (\ref{eq1}) with the hopping 
element (\ref{eq2}).
The resulting matrix is diagonalized
numerically using standard orthogonal decomposition methods.
It is important to notice that in contrast to the corresponding matrix of 
the Anderson model we do not obtain a sparsely occupied matrices. This 
requires more numerical effort and leads to a limitation of the
matrix size we can diagonalize to matrices not larger than $400\times 400$. 
As a result the distribution curves fluctuate stronger than in the case of 
nearest neighbor hopping models, where the matrices can be significantly
larger.

The level spacing distribution $P(s)$ of our model is analyzed and compared 
with the Poisson distribution 
and with the
distribution of the Gaussian orthogonal ensemble (GOE, $P^G(s)=\pi
\frac{s}{2} e^{-\pi \frac{s^2}{4}}$).
The choice of the GOE is due to the fact that our Hamiltonian obeys time
reversal symmetry.
As one can see in fig. \ref{fig1} the level-statistics for the quantum 
percolation regime does not follow the GOE
regime. In particular, for $s>2$ the distribution decays slower than
$P^G(s)$. This is in agreement with the prediction by Aronov et
al. \cite{aronov9495} for the situation near the Anderson type metal
insulator transition. It can be interpreted as the domination of the
statistics for larger level spacings by weakly overlapping states. 
According to our results this is not only a feature near the metal
insulator transition but it is present in the whole doping range.
In general the level repulsion of the quantum  percolation
model is weaker than the one for the GOE.
In order to investigate this behavior in more detail we analyze the
$\Delta_3$ statistics of the eigenvalue spectra. The latter is defined
for a different number $n$ of levels as \cite{metha91}
\begin{equation}
\Delta_3(n) = \frac{1}{n} {\mbox{Min} \atop A,B} \int_0^n (St(x') - A x' -
B)^2 d x' \; . 
\end{equation}
Here $St(x')$ is the staircase function. The result is shown in
fig. \ref{fig2}. The $\Delta _3$ statistics indicates the following
behavior for the 
doping dependence of the system: for low doping, far below the
classical percolation threshold, the system shows only weak level repulsion
and a tendency towards Poisson statistics. This is expected from common 
arguments because well-separated localized states are almost independently 
distributed, leading to a Poisson distribution.
For moderate doping, in the
vicinity of the classical percolation threshold, the level repulsion
increases and the system shows a tendency towards the Wigner statistics. 
This behavior indicates the beginning of the formation of overlaps between 
the states. As a result the eigenvalues experience level repulsion.
However, this tendency is reversed, when the doping concentration is further
increased above the percolation threshold. 
A possible explanation is a tendency towards uncorrelated $k$--space states 
of the fully doped (pure two--dimensional) system. 

To investigate the transition between spatially localized states and extended
states the sensitivity of the eigenvalues with respect to a change of the
boundary conditions is considered \cite{hatsugai93}. The Peierls substitution
of the hopping matrix elements
$ t \longrightarrow t e^{i \phi \Delta x} $ is used to vary continuously the boundary 
conditions in the Hamiltonian. Expanding the exponential function enables us 
to use perturbation theory
\cite{edwards72}
\begin{equation}
{\bf H} \longrightarrow {\bf H} + {\bf H_{\phi}}
\end{equation}
\begin{equation}
{\bf H_{\phi}} = \sum_{ij}(i t_{ij}(x_i-x_j) \phi c_i^{\dagger} c_j
 + h.c. ),
\end{equation}
where $x_j$ is the $x$-coordinate of site $j$.
Since ${\bf H_{\phi}}$ is purely imaginary the eigenvalues of the Hermitean 
Hamiltonian are affected only in second order perturbation theory
\begin{equation}
\label{thoulform}
\overline{\Delta E_M} = \overline{
\sum_{N \neq M} \frac{ | \langle \psi _M | {\bf H_{\phi}} | \psi _N \rangle | ^2}{
E_M - E_N }} \; .
\end{equation}
For the numerical calculation the average is taken over an ensemble of 
(typically 30) matrices. $N$ runs from $M-10$ to $M+10$ since the nearest 
energy levels contribute mainly to $\Delta E$. $ \overline{\Delta E_M} $ can 
be identified with the conductivity via the Kubo--Greenwood formula
\cite{thouless74}. Eqn. (\ref{thoulform}) is also known as the Thouless
formula for the conductivity.

A numerical investigation can only give information about 
localization lengths smaller or comparable with the system size
\cite{montambaux}.
The transition from localized to delocalized states, shown in fig.
\ref{fig4}, may indicate a real transition to delocalized states in the
infinite system. 
At least it will indicate a transition from exponentially to algebraically 
decaying states.
Furthermore, fig. \ref{fig4} shows that with increasing $\alpha$ the
crossover from localized to extended states gets more abrupt while the
absolute value of the energy shift decreases. $\alpha ^{-1}$, the 
characteristic length scale of the hopping processes, is always
much smaller than the system size.

In conclusion, we find 
a clear indication of a qualitative change of the system in
the level statistics as we go through the percolation threshold as shown in
figs. \ref{fig1},\ref{fig2}. This effect depends on the strength of doping
$c$ as well as on the range of the transfer $\alpha$.
To explain the onset of delocalization for the normal state of the high-T$_c$
cuprates with our model the diameter of the polaronic states should be 
chosen as $r_0 \approx 4 a$. 
This size is motivated by spatial inhomogeneities seen in experimental
observations (see e.g. inelastic neutron scattering data of 
ErBa$_2$Cu$_3$O$_x$ by Mesot et al. 
\cite{furrer} which are interpreted with similar cluster sizes).
Such polaronic states will give a percolation threshold for a doping
concentration of $c \approx $ 0.05. Therefore, the transition from
localized to delocalized states will occur near this concentration
if $r_0 \approx 4a$. \\[0.5cm]

M.L. thanks R.J. Gooding for stimulating discussions and for reading of the
manuscript.  



\vskip 2cm
{\Large \bf Figure captions} \\ \vskip 0.1cm
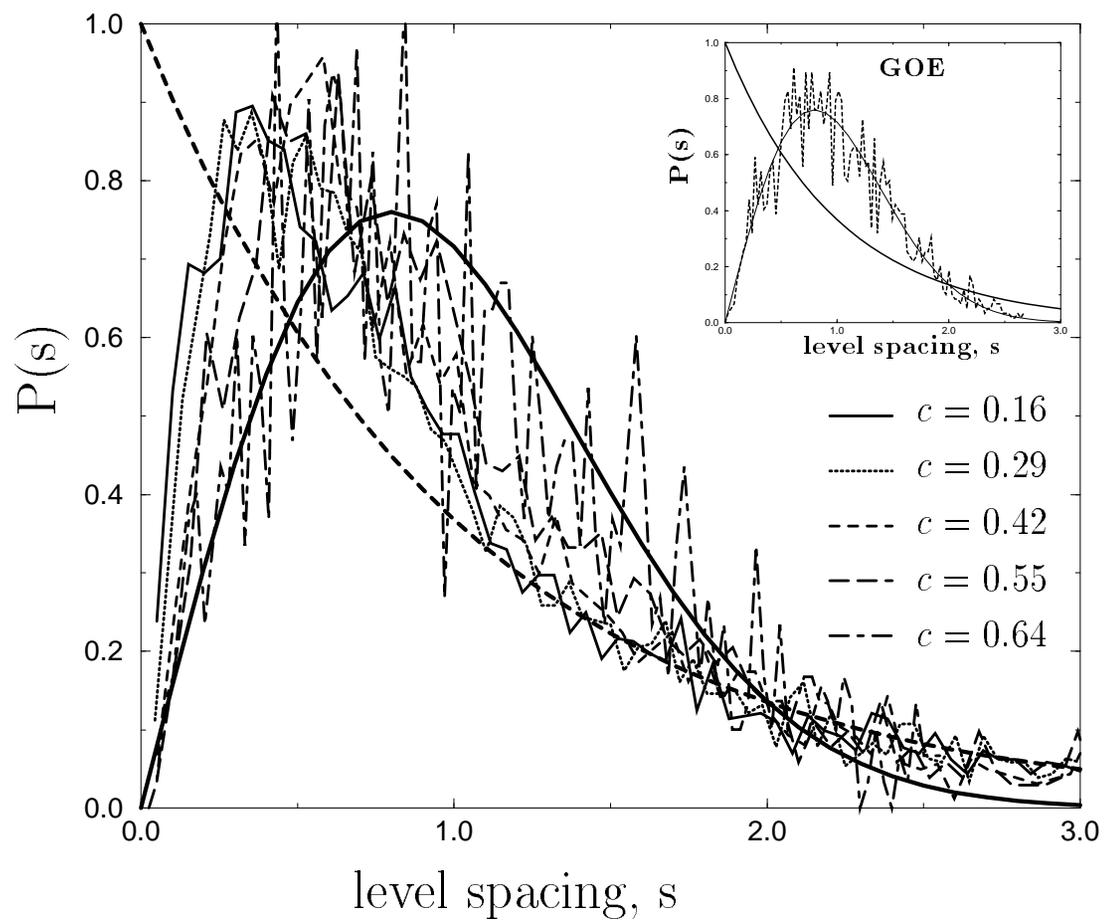
\begin{figure}
\caption{
For different concentrations, below and above the classical percolation
threshold, the level spacing distribution is drawn. It is compared with the
Wigner and the Poisson distribution. Note that for $s>2$ the distribution
decays slower than predicted by the Wigner distribution. 
The parameters for this plot are
$\alpha=0.5/a$ 
with the lattice constant $a$ and $ r_0=a$
For comparison the inset shows the result for the GOE ensemble of the same size. 
}
\label{fig1}
\end{figure}
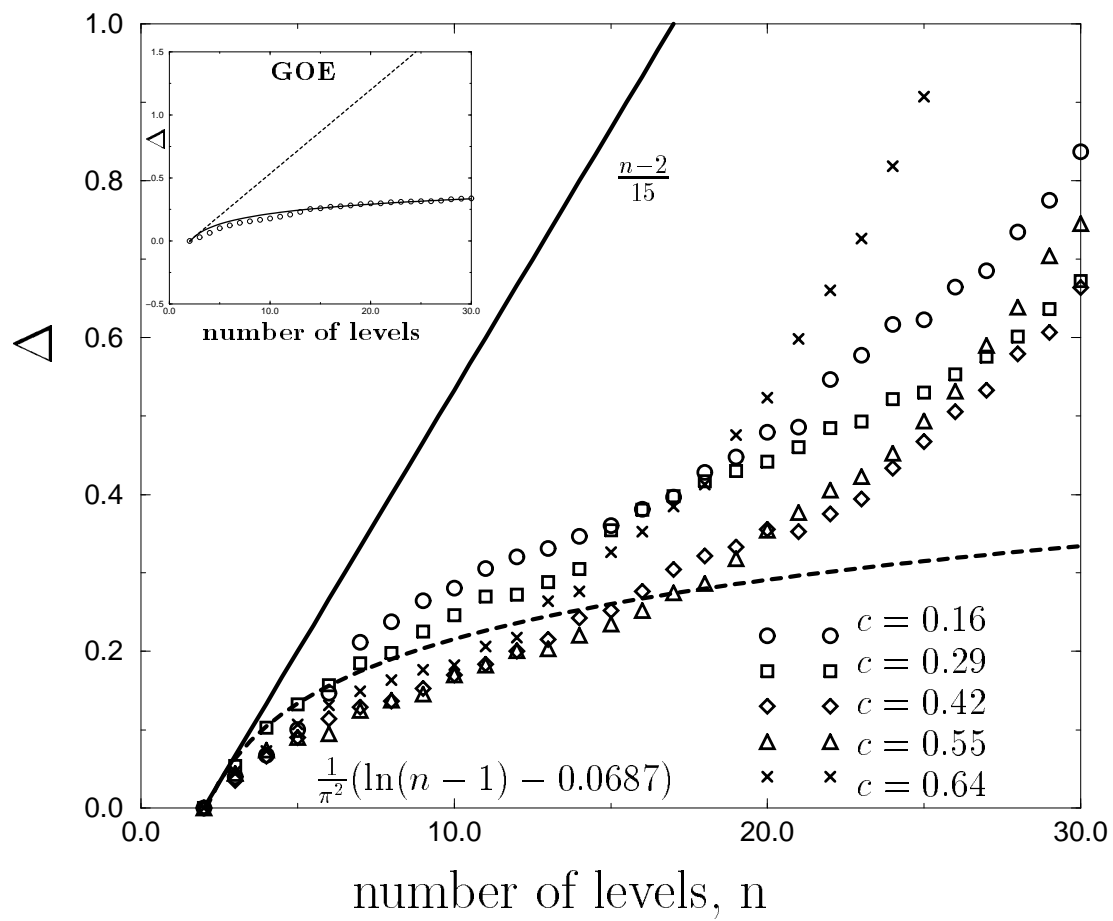
\begin{figure}
\caption{
For the same concentrations and parameters as in fig. 1 the results of the
calculated $\Delta _3$ statistic is plotted. The transition from a more 
Poisson like to a more Wigner like and back to a Poisson like distribution
can be clearly seen in 
this plot. The level repulsion is strongest near the classical percolation
threshold. Again for comparison the GOE ensemble is shown in the inset.
}
\label{fig2}
\end{figure}
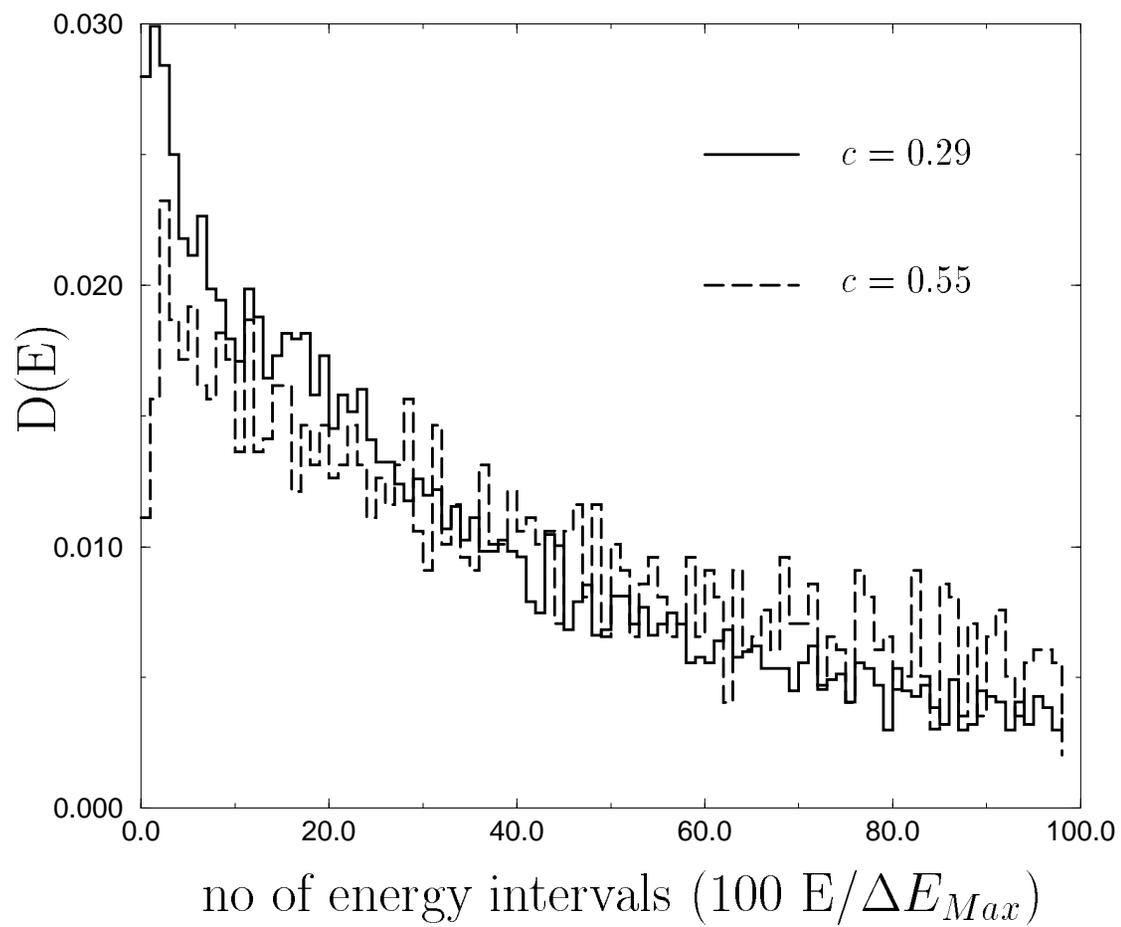
\begin{figure}
\caption{
The density of states for two different concentrations is drawn. The 
increase of the density of states towards small energy values is due to the
two-dimensionality of the system. Since $\alpha=0.5/a$ 
as in fig. 1 and in fig. 2 was chosen no peaks from parts
separated from the percolation backbone occur.
}
\label{fig3}
\end{figure}
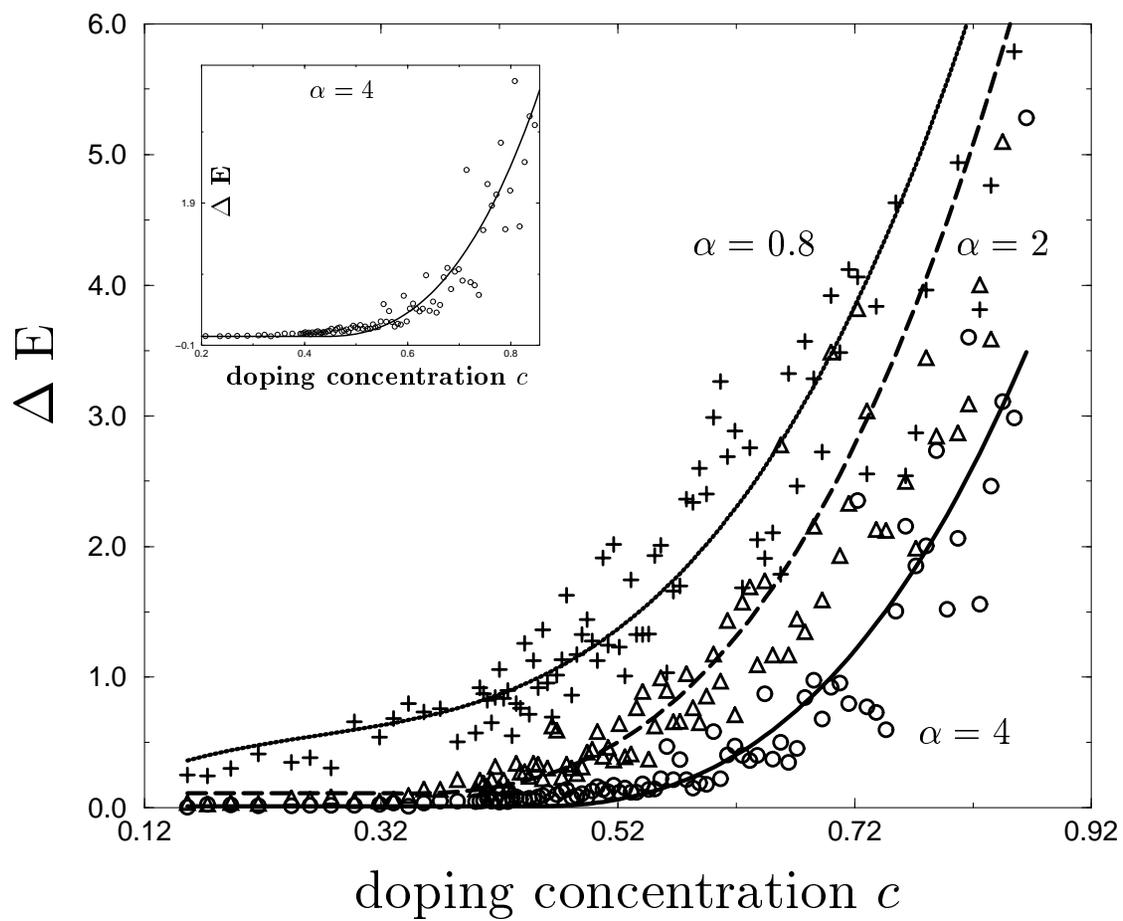
\begin{figure}
\caption{
For different values of $\alpha$ (in units of the inverse lattice constant) 
the sensitivity to the boundary conditions is 
shown in dependence of the doping concentration $c$. 
A
transition from localized to delocalized states can be seen
which for larger $\alpha$ gets more abrupt as can be seen in the inset. For 
this calculation $r_0$ is again $r_0=a$ 
}
\label{fig4}
\end{figure}

\end{document}